\documentclass[12pt]{elsarticle}
  
\usepackage{epsfig}
\usepackage{etoolbox}
\usepackage{graphicx}
\usepackage{amsmath}
\usepackage{amssymb}
\usepackage{amsfonts}
\usepackage[geometry]{ifsym}
\usepackage{moreverb}
\usepackage{url}
\usepackage[breaklinks]{hyperref}
\usepackage{xcolor}
\usepackage{subfig}
\usepackage{threeparttable}
\usepackage{verbatim}
\usepackage{epstopdf}
\usepackage{tikz}
\usepackage{todonotes}
\usepackage{amssymb}
\usepackage{epstopdf}
\usepackage{float}
\usepackage{algorithm}
\usepackage[noend]{algpseudocode}

\newcommand{\pcent}{{\%}}

\long\def\hide#1{}

\newcommand{\Cs}{\mathcal{C}}
\newcommand{\Ls}{\mathcal{L}}

\DeclareMathOperator*{\argmax}{arg\,max}
\makeatletter
\def\blfootnote{\xdef\@thefnmark{}\@footnotetext}
\makeatletter
\def\BState{\State\hskip-\ALG@thistlm}
\makeatother

\journal{IEEE ITSC 2017}
\title{\LARGE \bf
A Hidden Markov Model for Route and Destination Prediction 
}

\author{Yassine Lassoued\fnref{label}}
\author{Julien Monteil\fnref{label}}
\author{Yingqi Gu\fnref{labelUCD}}
\author{Giovanni Russo\fnref{label}}
\author{Robert Shorten\fnref{labelUCD}}
\author{Martin Mevissen\fnref{label}}
\fntext[label]{IBM Research - Ireland Lab, Control and Optimization Group. Address: Building $3$, IBM Technology Campus, Mulhuddart, Dublin, 15.}
\fntext[labelUCD]{University College Dublin - School of Electrical Engineering. Address: Belfield, Dublin, $4$.}

\begin{document}
\begin{frontmatter}

\begin{abstract}
We present a simple model and algorithm for predicting driver destinations and routes, based on the input of the latest road links visited as part of an ongoing trip. The algorithm may be used to predict any clusters previously observed in a driver's trip history. It assumes that the driver's historical trips are grouped into clusters sharing similar patterns. Given a new trip, the algorithm attempts to predict the cluster in which the trip belongs. The proposed algorithm has low temporal complexity. In addition, it does not require the transition and emission matrices of the Markov chain to be computed. Rather it relies on the frequencies of co-occurrences of road links and trip clusters. We validate the proposed algorithm against an experimental dataset. We discuss the success and convergence of the algorithm and show that our algorithm has a high prediction success rate.
\end{abstract}
\end{frontmatter}

\section{Introduction}
\label{sec:intro}

\hide{Vehicles are undergoing a transformation as they are transitioning from being isolated entities to becoming connected devices~\cite{Connectivity}. The recent study~\cite{McKinseyInternet}, for example, showed that drivers are not anymore willing to buy cars without connectivity capabilities. This led to a shift in research and development, in both industry and academia, towards the development of personalised services for drivers, leveraging the connectivity capabilities offered by modern vehicles. Examples of such services include speed advisory systems, intended to recommend optimised speeds to drivers for a given cost function for a network of vehicles~\cite{Gri_Rus_Sho_16}, parsing engines, which can be used to implement traffic balancing systems~\cite{7795769}, engine management systems for plug-in electric vehicles to tailor the usage of electric batteries based on the driver's behaviour~\cite{Cri_16}.}

As personalised driving experience expectations are growing~(\cite{ibm-automotive-2025}), so is the need to automate the prediction of the driver's behaviour. Understanding the driver's intentions is a prerequisite to enabling personalised assistance functions, such as personalised risk assessment and mitigation, speed advice~(\cite{Gri_Rus_Sho_16, juliengiovanni, 7795777}), rerouting~(\cite{7795769}) infrastructure systems~(\cite{victorjulien}), engine management systems~(\cite{Cri_16}), etc. Amongst the key driver intentions to predict, destination and route have been the subject of several research efforts, see for example~(\cite{1706730,6544830,Che_10,2008-01-0201,6728224}) and the references therein.

\hide{People's daily trips seem to follow regular patterns, both spatially and temporally.
A recent study \cite{Barabasi} suggests that the level of predictability of routes and destinations for drivers is above $90\pcent$. A possible explanation is that drivers tend to follow the same routes in their routine trips. Even when better (i.e., faster and/or shorter) routes exist, drivers seem to tend to stick to what they are familiar with. This observation contributed to motivate, over the past few years, research efforts devoted to the development of route and destination prediction algorithms, see for example~(\cite{1706730,6544830,Che_10,2008-01-0201,6728224}) and the references therein.}

While this paper is inspired by the recent numerous activities on route and destination prediction, it offers a number of novelties. In particular, we present here a new destination and route prediction algorithm with minimal complexity. The proposed algorithm predicts the route and destination of a driver given the latest road links visited as part of the ongoing trip. The algorithm may be used to predict patterns previously observed as part of a driver's trip history. It simply assumes that the driver's historical trips are grouped into clusters sharing similar patterns (e.g., clustering by destination, by route similarity, etc.). Given a new trip, the algorithm attempts to predict the cluster within which this trip belongs.

The complexity of our algorithm is bi-linear in the number of the input visited road links, and the total number of clusters. We evaluated our algorithm against a synthetic dataset of trips of a single fictional driver, covering a one-year period. \hide{For this dataset, we show that our algorithm always ends up predicting the correct route/destination given a sufficient number of observations.} We report on the convergence rate of our algorithm, expressed as the number of road links required to correctly predict the route/destination. To the best of our knowledge, this performance indicator tends to be neglected in the literature.

\hide{Finally, we also discuss the trade-off between accuracy of the predicted destination and the rate of convergence of the algorithm. In particular, we provide guidelines on how certain parameters of the algorithm can be tuned to improve predictions.} 
\section{Proposed Approach}
\label{sec:approach}
\begin{figure}[H]
\begin{center}
\includegraphics[width=0.75\textwidth]{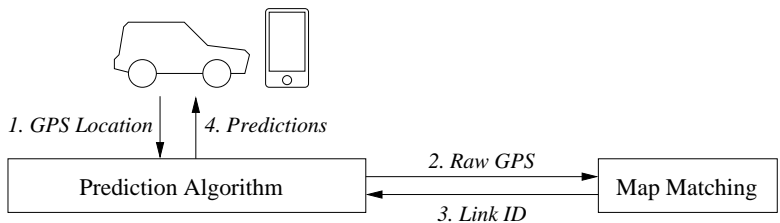}
\caption{\label{fig:architecture}Destination/Route Prediction Workflow}
\end{center}
\end{figure}

The overall principle of the destination/route prediction approach we propose is illustrated in Figure~\ref{fig:architecture}. A registered vehicle or driver needs to be equipped with a client application in order to interact with the predictor. This is typically a mobile phone app. The client periodically sends the car probe data to the destination/route predictor, and receives predictions based on these. The predictor is trained on the driver's historical data. At the end of a new trip, the trip data are added to the driver's historical database, allowing the predictor to update its prediction models.

The prediction algorithm requires that GPS positions be mapped to road links, as trajectories are modelled as sequences of road links rather than GPS coordinates. Therefore a map matching component is needed.

\subsection{Data Gathering and Pre-Processing}
Raw trip data for a given driver and vehicle are initially gathered by the client application in the form of sequences of GPS positions with time stamps. Trips with multiple destinations are split into multiple trips. We assume that the client application is responsible for detecting the start and end of a trip. This may be achieved in several more or less sophisticated ways. In modern cars, trip data may be available through the OBD2 interface, which may be used as a primary way to detect the start and end of trips. This may be further refined based on the duration of stops, whether the engine was on or off, distance to road, etc. Trip detection is beyond the scope of this article, but is a prerequisite to our algorithm and assumed to be the responsibility of the client.

\begin{figure}[H]
  \begin{center}
  \input{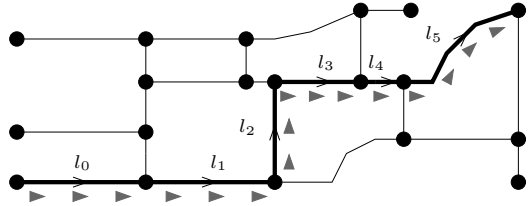}
  \caption{\label{fig:snap}Map Matching Example}
	\end{center}
\end{figure}

As mentioned earlier, our prediction algorithm requires trips in the form of sequences of road links rather than GPS coordinates. Therefore raw trip data collected by the client are mapped to directed road links using a map matching component. Figure~\ref{fig:snap} shows an example of a car trajectory, represented by a sequence of oriented arrow heads, mapped to a sequence of links $[l_0, ..., l_5]$ (bold lines). Several consecutive raw positions may map to the same road link, but each link is considered only once in a sequence.

\subsection{Clustering}
\label{sec:cluster}

\begin{figure}[H]
  \begin{center}
  \input{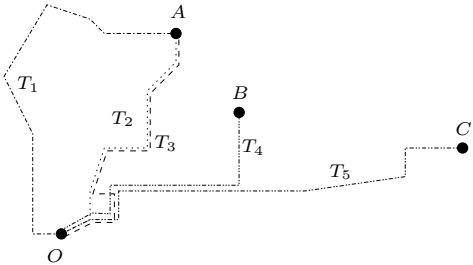}
  \caption{\label{fig:cluster} Clustering trips $\{T_1, ..., T_5\}$ by destination (assuming that $O$ is the origin) leads to three destination patterns $\{T_1, T_2, T_3\}, \{T_4\}, \{T_5\}$, while clustering by route similarity leads to four route patterns $\{T_1\}, \{T_2, T_3\}, \{T_4\}, \{T_5\}$.}
\end{center}
\end{figure}

In reality, our prediction algorithm is not limited to predicting destinations or routes. Rather, it can be trained to predict any patterns or clusters. The idea is that you provide it with training data in the form of clustered historical trips, where each trip is assigned a cluster, usually representing a pattern. For example historical trips may initially be clustered by destination, in which case the algorithm will predict destinations, or by route similarity, in which case it will predict route patterns. In all cases, we assume that a clustering step \hide{using an adequate clustering scheme} is required prior to the training, and that the algorithm is used to predict the cluster of a trip. Here we discuss two types of clustering of trips: clustering by destination and clustering by route, as illustrated in Figure~\ref{fig:cluster}.

\paragraph{Clustering by Destination} In this case we cluster trips based on the proximity between the GPS coordinates of their destinations. We rely on traditional clustering techniques, such as hierarchical clustering or k-means clustering~\cite{hastie2009unsupervised}. In the former, the key clustering parameter is a distance threshold above which a cluster node is to be further split. In the latter, the key parameter is the number of clusters, which may be determined  given a relative variance threshold. 
\paragraph{Clustering by Route Similarity} In this case we cluster trips based on a similarity measure between their routes. We define the similarity of two routes as the ratio of number of shared road links and the total number of links in both routes. We then compute the distance matrix between all the trips and apply a hierarchical clustering choosing a dissimilarity of 0.2 as the threshold.

\subsection{Training}
\label{sec:train}

For a given vehicle and driver, we train our prediction algorithm using pre-processed and clustered historical trips. Assuming that all trips have been assigned to clusters, we define the set of clusters $\Cs=\{1,\ldots, n\}$ of cardinal $|\Cs| = n$. Consider $\Ls=\{1,\ldots, m\}$ the set of all road links belonging to all the historical trips, where $|\Ls| = m$. Each trip is defined as a sequence of one or more links in $\Ls$, with their time stamps, and is associated with a cluster or pattern in $\Cs$. 

The training simply consists in computing the link-cluster co-occurrence frequency matrix $F$ of dimension $m \times n$, where $F_{i, j}$ is the frequency of cluster $j \in \Cs$ in trips containing link $i \in \Ls$. From this matrix we can infer the following probabilities required by our prediction algorithm.

The probability to be on link $l$ knowing that the current trip belongs to cluster $C=c$:
\begin{equation}
p(l|C=c)=\frac{\text{\# trips traversing $l$ in cluster $c$}}{\text{\# trips in cluster $c$}} =  \frac{F_{l, c}}{\sum \limits_{i=1}^m F_{i, c}}.\label{eq:00}
\end{equation}

The probability to be in cluster $C=c$ knowing that the current link is $l$:
\begin{equation}
p(C=c|l)=\frac{\text{\# trips traversing $l$ in cluster $c$}}{\text{\# trips passing via $l$}} = \frac{F_{l, c}}{\sum \limits_{j=1}^n F_{l, j}}.\label{eq:01}
\end{equation}


\subsection{Prediction}
\label{sec:predict}

The prediction is based on the provision of a sequence of $k$ road links $l_{1:k} = (l_1, ..., l_k)$ visited as part of the ongoing trip, for which we want to predict the cluster (route or destination). The cluster prediction algorithm relies on the Markovian formalism, where clusters act as hidden states and links as observations. The time step of the process is the time needed for the vehicle to enter a different road link (observation). For time $t=k$, $C_k$ denotes the hidden cluster state that we want to estimate, and $l_k$  the current link observation. According to the Markovian formalism, the probability of $C_{k+1}$ given $l_{1:k}$ is provided by:

\begin{equation}p(C_{k+1}=c|l_{1:k})=  \sum \limits_{i=1}^n p(C_{k+1}=c|C_k=i) p(C_{k}=i|l_{1:k}).\label{eq:02}\end{equation}

Probability $p(C_{k+1}|C_k)$ is given by the transition matrix of the hidden Markov model. We assume that a driver will not change his destination or route on the fly. Therefore, the hidden cluster state is considered independent of the time step, and the transition matrix is simply the identity matrix. Therefore, the propagation model of \eqref{eq:02} becomes:

\begin{equation}p(C_{k+1}|l_{1:k})=p(C_{k}|l_{1:k}).\label{eq:03}\end{equation}

This reduces the model to Naive Bayes. According to Bayes' theorem, we obtain the following recursive formula:

\begin{equation}p(C|l_{1:k},l_{k+1})\propto p(l_{k+1}|C)p(C|l_{1:k}).\label{eq:04}\end{equation}

The prior probability $p(C|l_1)$ is provided by equation \eqref{eq:01} above. As can be seen, the probability $p(l_{k+1}|C)$ does not need to be estimated using machine learning techniques such as the Baum-Welch algorithm~\cite{baum1970maximization}. Instead, it only relies on counting the trip occurrences, as presented in Section~\ref{sec:train}, equations~\eqref{eq:00} and~\eqref{eq:01}.

It is possible that a trip includes a wrong link observation, due to noisy GPS observations or erroneous map matching, or that the current link $l_{k+1}$ has never associated with cluster $C$. In such a case, equation~\eqref{eq:04} will result in $p(l_{k+1}|C) = 0$, leading to zero probability for cluster $C$ for all the subsequent steps. The risk is that a probability 0 may be associated with the correct cluster. To avoid this problem, we introduce a small constant r, inspired from the Google PageRank algorithm~\cite{page1999pagerank}, and correct the probability of equation~\eqref{eq:04} as follows:

\begin{equation}
p(C|l_{1:k+1}) \leftarrow \frac{r}{n} + (1-r)p(C|l_{1:k+1}). \label{eq:05}
\end{equation}

It is important to note though that while the introduction of such a constant is meant to help avoid probability 0 being associated with the correct cluster, it comes at the expense of slowing down the convergence rate. The higher $r$ is, the slower the algorithm will converge. An optimum value for $a$ may be learned via traditional machine learning techniques.

The resulting algorithm is straightforward. The only defined parameter is the tolerance threshold, defined as $\epsilon$, that is used as a termination criterion.
\begin{algorithm}[H] 
\caption{Prediction with clusters as hidden states}\label{algo:prediction}
\begin{algorithmic}[1]
\Procedure{ClusterPrediction}{}
\BState \emph{Initialisation}:
\State $\epsilon \gets \text{tolerance {\em // (e.g., 0.01)}}$
\State $n \gets |C| \ \text{{\it // total number of clusters}}$
\State $k \gets 1$
\For{$j\in\mathcal{C}$}
\State $P_j \gets p(C=j|l_1)$
\EndFor
\BState \emph{Algorithm core}:
\While {$\max_{j \in [1:n]} P_j < 1 - \epsilon$} 
\State $\text{Wait for new observed link}:\ l_{k+1}$
\For{$i\in\mathcal{C}$}
\State $P_j \gets P_j\cdot p(l_{k+1}|C=j)$
\EndFor
\State $\text{normalise } (P_j)_{j \in [1:n]}$
\For{$i\in\mathcal{C}$}
\State $P_j \gets r/n+(1-r)P_j$
\EndFor
\State $k \gets k+1$
\EndWhile
\BState \emph{Final prediction}:
\State $\text{Predicted cluster} \gets \argmax_{j \in [1:n]} P_j$
\EndProcedure
\end{algorithmic}
\end{algorithm} 

Note that the consideration of context (day and time of the trip) simply requires to compute frequencies conditionally to the context. However, the selection of driver-tailored contexts can be a challenging task, as discussed in the following section.

\subsection{Context Considerations}
\label{sec:context}

Here we discuss the introduction of context in our prediction algorithm. Context is represented here as a variable $\gamma$ belonging in a {\em multi-dimensional discrete} context space~$\Gamma$. It can be thought of as a cell in a multi-dimensional matrix or hypercube.
For example, since people's activities (and therefore destinations and routes) mainly depend on time, we can use trip timestamps as temporal context. We define temporal context as a two-dimensional variable $\gamma = (d, t)$, where  $d \in [0:6]$ is a day of the week, and $t$ is a time of the day, discretised into time bins (e.g., twenty-minute or one-hour bins). Each link $l_i, i \in [1:k]$, has an associated time stamp (e.g., time link was entered), and therefore a temporal context $\gamma_i = (d_i, t_i)$. This can be thought of a cell in the two dimensional matrix illustrated in Figure~\ref{fig:context}.

We extend the above prediction algorithm with context by considering observations (visited links) in their context. This means that, in the training phase, instead of computing one single frequency matrix $F$, we compute multiple matrices $F^\gamma$, one for each discrete context value $\gamma$ in the context space $\Gamma$. We revise the probability expressions \eqref{eq:05} as follows:\\
The probability to be on link $l$ knowing that the current trip belongs to cluster $C=c$ and that context is $\gamma$:
\begin{equation}
p(l|C=c, \gamma) = \frac{F_{l, c}^\gamma}{\sum \limits_{i=1}^m  F_{i, c}^\gamma}.\label{eq:06}
\end{equation}
The probability to be in cluster $C=c$ knowing that the current link is $l$ and that context is $\gamma$:
\begin{equation}
p(C=c|l, \gamma) = \frac{F_{l, c}^\gamma}{\sum \limits_{j=1}^n F_{l, j}^\gamma}.\label{eq:07}
\end{equation} 

In the prediction phase, given an observed road link $l$ with context $\gamma \in \Gamma$, we retrieve from the prediction model the pattern probability distribution for the same link in the same context (cell). While this is meant to help improve precision, it may turn out to be problematic due to historical data sparsity and variability. Suppose we represent context by day of the week and one-hour time bins of the day. Assume, for example, that you always drive to work sometime between eight and nine o'clock in the morning, but that today, Wednesday, you are two hours late. For a given road link $l$ visited during your trip to work, no historical observation for the same link $l$ and for the same temporal context (Wednesdays between ten and eleven o'clock) will be found as illustrated in figure~\ref{fig:context}. However, we may have historical observations for $l$ in different contexts (greyed cells). Given that it is Wednesday morning, common sense suggests that you are likely to be driving to work. There is a need then to allow for some flexibility when matching the context of the current observation (link) to those of historical observations.
 
\begin{figure}[H]
  \begin{center}
  \input{xfig-context.pstex_t}
  \caption{\label{fig:context}Context Matching: Link $l$ is observed in the current trip for the first time on a Wednesday between 10 and 11 o'clock, but has been observed in historical trips at other times (grey cells).}
	\end{center}
\end{figure}

\begin{figure}[H]
  \begin{center}
  \includegraphics[width=0.6\textwidth]{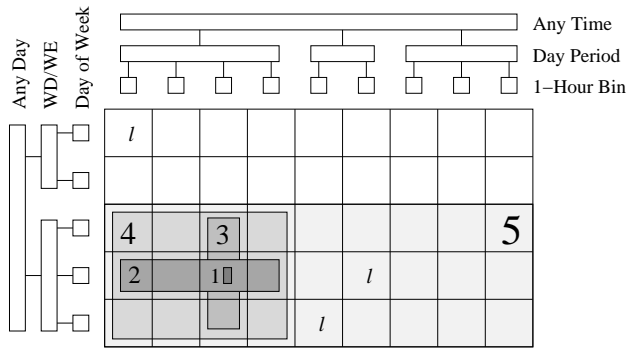}
  \caption{\label{fig:expansion}Context Expansion Example}
	\end{center}
\end{figure}

We propose an automatic context expansion method allowing for flexible context matching. In this solution, each context dimension is represented as a hierarchy of bins with varying (increasing) granularity. Figure \ref{fig:expansion} shows an example where days of the week are grouped initially into working days and week-end, then refined into actual week days. In the same way, time of the day may be represented by broad day period bins (morning, afternoon, evening, etc.), then refined into narrower bins (down to hourly, or twenty-minute bins). A context matcher allows us to start with the finest context cell in the context hypercube, then expand according to one or multiple context dimensions, until a minimum number of observations is reached. The context expansion method uses an expansion strategy which defines the order and magnitude of expansion according to each axis. For instance, you may want to expand the time of the day bin to the day period, before attempting to expand the context along the days of the week axis, and so on.

\hide{
\subsubsection{Context Similarity Weighting}
In this solution we rely on a similarity measure $\sigma$ between context values, such that $\sigma(\gamma, \gamma') \in [0, 1]$ for $\gamma, \gamma' \in \Gamma$. Given an observed link $l$, in a given context $\gamma$, we consider all the observations of $l$ in all contexts, and we redefine the probabilities expressed in formulae \eqref{eq:06} and \eqref{eq:07} as follows:\\
\begin{equation}
p(l|C=c, \gamma) = \frac{1}{\sum\limits_{\gamma' \in \Gamma} \sigma(\gamma, \gamma')} \sum\limits_{\gamma' \in \Gamma} \sigma(\gamma, \gamma') \frac{F_{l, c}^{\gamma'}}{\sum \limits_{i=1}^M F_{i, c}^{\gamma'}}\label{eq:08}
\end{equation}
\begin{equation}
p(C=c|l, \gamma) = \frac{1}{\sum\limits_{\gamma' \in \Gamma} \sigma(\gamma, \gamma')} \sum\limits_{\gamma' \in \Gamma} \sigma(\gamma, \gamma') \frac{F_{l, c}^{\gamma'}}{\sum \limits_{j=1}^N F_{l, j}^{\gamma'}}\label{eq:09}
\end{equation}
}

\section{Experimental Results}
\label{sec:results}

\subsection{Dataset}

The dataset was generated in the following way. We selected an area of interest in Dublin city from OpenStreetMap, with 7 locations as origins/destinations (``home", ``friend's home", ``workplace", ``city centre", ``shopping centre", ``hospital", ``childcare") representing the usual destinations of a fictional driver. This led to 21 origin/destination pairs. For each pair, we generated up to 3 routes representing different options a driver may choose from. We then fed these into SUMO (Simulation of Urban MObility)~(\cite{SUMO2012}) to generate GPS trajectories of simulated cars. These GPS tracks were generated in a probabilistic manner following a weekday/weekend scenario. For example, during weekdays the driver has a higher probability (95\%) to drive from home to work each morning (9am - 10am), while during weekends the driver has higher probability (70\%) to drive to the city centre from home. Using the resulting GPS tracks as input, we then added random noise to the GPS coordinates, by sampling from a uniform distribution around the input GPS points, with a maximum bias of 10m. Such an error can be seen as a worst case scenario for GPS noise. The resulting noisy GPS tracks were then stored as the raw GPS data for the driver historical trips.

As our prediction algorithm expects road links rather than GPS coordinates, we mapped the raw noisy GPS tracks back to road links using IBM Streams. This resulted in trips in the form of sequences of road links together with their timestamps. We clustered the so-obtained trips by destination and by route similarity as described in section~\ref{sec:cluster}, leading to 7 destination clusters and 34 route clusters. Finally we split the trips ($50\pcent$-$50\pcent$) into train and test datasets and computed the frequency matrices from the train dataset as described in section~\ref{sec:train}.

\subsection{Destination Prediction Results}
\begin{figure}[H]
\begin{center}
\begin{tabular}{cc} 
\subfloat[]{\includegraphics[width=0.75\textwidth]{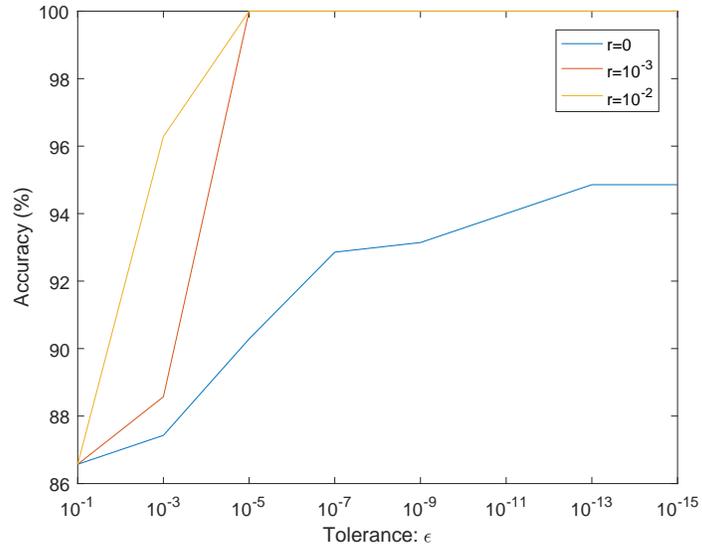}\label{dest00}}\\
\subfloat[]{\includegraphics[width=0.75\textwidth]{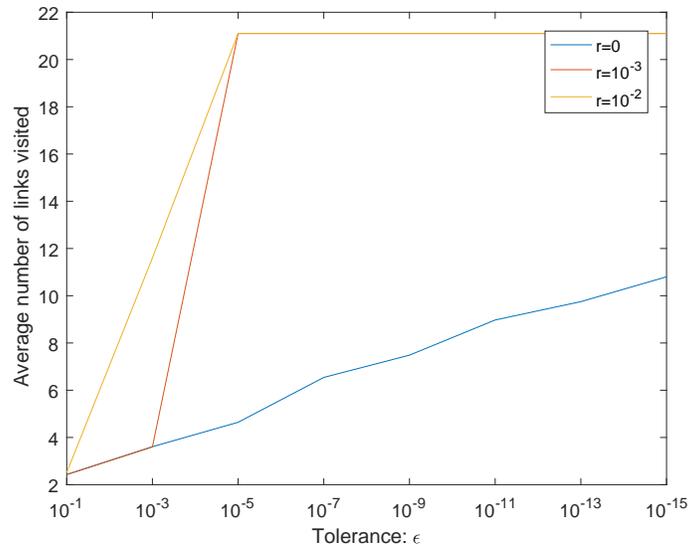}\label{dest01}}
\end{tabular}
\caption{Destination prediction accuracy levels and their required number of links travelled, as a function of tolerance $\epsilon$, and for various values of the page rank constant $r$.}
\label{dest02}
\end{center}
\end{figure}

Figure~\ref{dest02} shows the results of our algorithm for destination prediction for various values of the page rank constant and the prediction tolerance (cf. Algorithm~\ref{algo:prediction}). With the page rank constant $r$ set to 0, we observe a slow convergence to an accuracy of $94.8\%$ until a plateau is reached as the tolerance threshold increases. This confirms that the proposed algorithm is very competitive in comparison to other approaches in the literature, see~(\cite{1706730,6544830}) for instance, with prediction accuracy levels higher than $90\%$. The average number of links needed to predict a correct destination with $94.8\%$ accuracy is 9.8, which seems to be a good result given that the average number of links per trip is 87.7 for the considered dataset. This means that correct destinations are predicted $94.8\%$ of the time when the driver has been through about $10\%$ of his trip.

Conforming to equation~\eqref{eq:04}, prediction errors can be avoided using a nonzero page rank constant $r$. However, this comes at the expense of a higher number of links needed to correctly predict destination, as can be seen in Figure~\ref{dest01}. The average number of links needed to correctly predict destinations jumps to 21.1 with $r = 10^{-2}$. This corresponds to about a quarter of the whole trip in average, which may not be sufficient for some applications, such as risk mitigation applications requiring that the driver destination be predicted at an early stage of the trip.

\subsection{Route Prediction Results}
\begin{figure}[H]
\begin{center}
\begin{tabular}{cc} 
\subfloat[]{\includegraphics[width=0.75\textwidth]{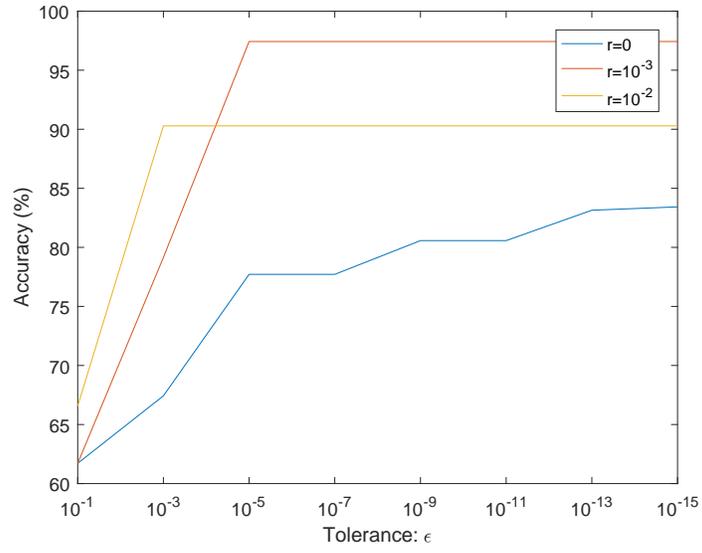}\label{dest03}}\\
\subfloat[]{\includegraphics[width=0.75\textwidth]{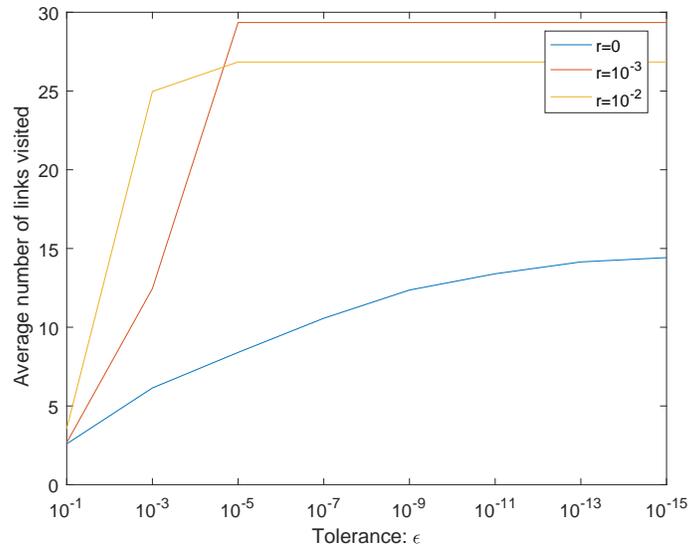}\label{dest04}}
\end{tabular}
\caption{Route prediction accuracy levels and their required number of links travelled, as a function of tolerance $\epsilon$, and for various values of the page rank constant $r$.}
\label{dest05}
\end{center}
\end{figure}

Figure~\ref{dest05} shows the results of our algorithm for route prediction for various values of the page rank constant and the prediction tolerance (cf. Algorithm~\ref{algo:prediction}). With the page rank constant $r$ set to 0, we observe a slow convergence to an accuracy of $84.0\%$ until a plateau is reached. The average number of links needed to predict the correct route with this level of accuracy is 14.1, which is satisfactory as it corresponds to approximately $15\%$ of the trip.

When setting the constant $r$ to a nonzero value, we observe an increase in the accuracy levels, reaching 97.4\% for $r=10^{-2}$. Again, this comes at the expense of a higher number of links needed to correctly predict the route, which jumps to 29.2, as can be seen in Figure~\ref{dest04}. This corresponds to more than a third of the trip in average, which again may not be sufficient for some applications. A solution to improve the convergence rate, other than optimising parameter $r$, may be to add a pre-processing step after the map-matching algorithm to remove the occurrences of jitters and stubs, as suggested in~\cite{1706730}.

\subsection{Accuracy as a Function of the Number of Visited Links}

\begin{figure}[H]
  \begin{center}
  \includegraphics[width=0.75\textwidth]{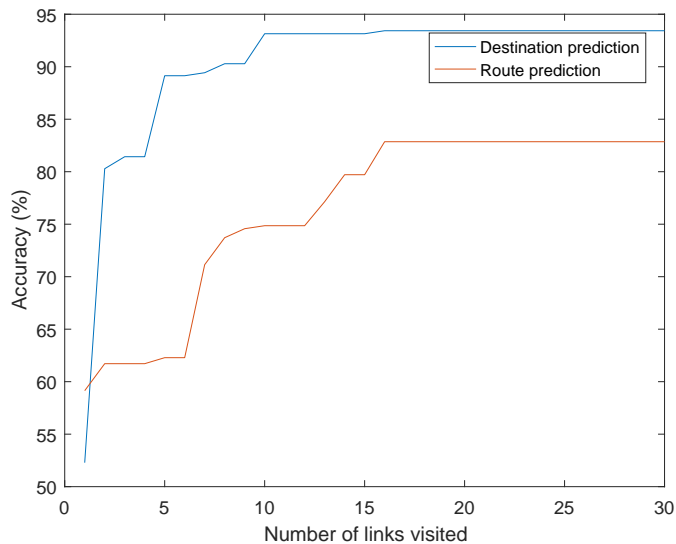}
  \caption{Prediction accuracy levels as a function of the number of links visited.}
	\label{lin}
	\end{center}
\end{figure}

Figure~\ref{lin} shows the prediction accuracy rate as a function of the number of links visited for both destination and route prediction. This is a rarely discussed evaluation metric in the literature. The idea is to run the prediction algorithm against the test dataset, and record the percentage of correct predictions as a function of the number of visited links. In this experiment, a prediction is considered as correct if the predicted destination/route with the highest probability corresponds to the actual destination/route of the trip.

Unsurprisingly, and in light of the previous figures~\ref{dest02} and~\ref{dest05}, the accuracy rate increases until it reaches a plateau. This plateau corresponds to 93.1\% after 10 links for destination prediction, and 83.3\% after 16 links for route prediction.
\hide{Note that the prediction accuracy results are slightly worse than those of the previous figures, as the termination condition of Algorithm~\ref{algo:prediction} is now different.}

\subsection{On the Comparison with Existing Algorithms}

\hide{We mentioned that several route and destination prediction algorithms have been proposed in the literature.}
While we leave a more thorough case by case comparison for future work, we discuss here several properties that need to be taken into consideration when comparing existing approaches.
\begin{itemize}
\item {\bf Pre-processing Requirements:} Prediction algorithms may rely on GPS trajectories only~(\cite{2008-01-0201,Che_10}), on map matching (the case of our algorithm), or on map matching followed by a correction step, e.g., the filtering of jitters and stubs~(\cite{1706730}).
\item {\bf Output of the Algorithm:} Algorithms may predict destination~(\cite{6728224}), route or/and destination~\cite{1706730}, or, more generally, clusters.
\item {\bf Complexity of the Algorithm:} Prediction algorithms may require trajectory mining computations~(\cite{6544830, Che_10}), or may be less demanding if relying on Markovian assumptions~(\cite{1706730}), but may rely on a map matching algorithm. In our case the complexity of the prediction algorithm itself is $O(\text{\#visited links} \times \text{\#clusters})$.
\item {\bf Convergence of the Accuracy Rate:} This is a rarely discussed performance indicator for destination/route prediction algorithms, however it is important to understand how the prediction accuracy of the algorithm evolves with the trip (e.g., prediction accuracy as a function of the number of links, or proportion of trip).
\item{\bf Train vs. Test Data:} The proportion of trained vs. tested dataset, which is 50\% in our case, may vary across works and affect the accuracy results. 
\item {\bf Benchmark Data:} The datasets against which an algorithm is benchmarked are important as their predictability levels may vary. In our case, we generated a synthetic dataset, which means that all the trips are predictable or, to reuse the formulation of~(\cite{2008-01-0201}), belong to the ``repeated trip" category.
\end{itemize}

\section{Conclusion}
\label{sec:conclusion}

In this paper we presented a simple yet novel algorithm for route and destination prediction, based on the sequence of visited road links as input, and on the formalism of Hidden Markov Models as a method. We tested our algorithm using a synthetic dataset, and showed that it leads to high prediction accuracy and convergence rates for both destination and route prediction. Extension of this work to community-shared vehicle, such as vehicles shared by a family or a company, will be the subject of our future work.

\section*{Acknowledgement}
This work has been conducted within the ENABLE-S3 project that has received funding from the ECSEL Joint Undertaking under grant agreement no 692455. This joint undertaking receives support from the European Union Horizon 2020 research and innovation programme and Austria, Denmark, Germany, Finland, Czech Republic, Italy, Spain, Portugal, Poland, Ireland, Belgium, France, Netherlands, United Kingdom, Slovakia, Norway.

\section*{References}
\bibliographystyle{elsarticle-harv} 
\bibliography{ref}

\end{document}